\documentstyle[12pt]{article}

\catcode`\@=11
\long\def\@makefntext#1{
\protect\noindent \hbox to 3.2pt {\hskip-.9pt
$^{{\ninerm\@thefnmark}}$\hfil}#1\hfill}		

\def\@makefnmark{\hbox to 0pt{$^{\@thefnmark}$\hss}}  

\def\ps@myheadings{\let\@mkboth\@gobbletwo
\def\@oddhead{\hbox{}
\rightmark\hfil\ninerm\thepage}
\def\@oddfoot{}\def\@evenhead{\ninerm\thepage\hfil
\leftmark\hbox{}}\def\@evenfoot{}
\def\sectionmark##1{}\def\subsectionmark##1{}}

\renewcommand{\thefootnote}{\fnsymbol{footnote}}

\newcounter{sectionc}\newcounter{subsectionc}\newcounter{subsubsectionc}
\renewcommand{\section}[1] {\vspace*{0.6cm}\addtocounter{sectionc}{1}
\setcounter{subsectionc}{0}\setcounter{subsubsectionc}{0}\noindent
	{\normalsize\bf\thesectionc. #1}\par\vspace*{0.4cm}}
\renewcommand{\subsection}[1] {\vspace*{0.6cm}\addtocounter{subsectionc}{1}
	\setcounter{subsubsectionc}{0}\noindent
	{\normalsize\it\thesectionc.\thesubsectionc. #1}\par\vspace*{0.4cm}}
\renewcommand{\subsubsection}[1]
{\vspace*{0.6cm}\addtocounter{subsubsectionc}{1}
	\noindent {\normalsize\rm\thesectionc.\thesubsectionc.\thesubsubsectionc.
	#1}\par\vspace*{0.4cm}}

\newcounter{appendixc}
\newcounter{subappendixc}[appendixc]
\newcounter{subsubappendixc}[subappendixc]

\renewcommand{\appendix}[1] {\vspace*{0.6cm}
        \refstepcounter{appendixc}
        \setcounter{figure}{0}
        \setcounter{table}{0}
        \setcounter{equation}{0}
        \renewcommand{\thefigure}{\Alph{appendixc}.\arabic{figure}}
        \renewcommand{\thetable}{\Alph{appendixc}.\arabic{table}}
        \renewcommand{\theappendixc}{\Alph{appendixc}}
        \renewcommand{\theequation}{\Alph{appendixc}.\arabic{equation}}
        \noindent{\bf Appendix \theappendixc #1}\par\vspace*{0.4cm}}

\def\abstracts#1{{

\centering{\begin{minipage}{12.2truecm}\footnotesize\baselineskip=12pt\noindent
	\centerline{\footnotesize ABSTRACT}\vspace*{0.3cm}
	\parindent=0pt #1
	\end{minipage}}\par}}

\renewenvironment{thebibliography}[1]
	{\begin{list}{\arabic{enumi}.}
	{\usecounter{enumi}\setlength{\parsep}{0pt}
\setlength{\leftmargin 1.25cm}{\rightmargin 0pt}
	 \setlength{\itemsep}{0pt} \settowidth
	{\labelwidth}{#1.}\sloppy}}{\end{list}}

\newcounter{itemlistc}
\newcounter{romanlistc}
\newcounter{alphlistc}
\newcounter{arabiclistc}

\newcommand{\fcaption}[1]{
        \refstepcounter{figure}
        \setbox\@tempboxa = \hbox{\footnotesize Fig.~\thefigure. #1}
        \ifdim \wd\@tempboxa > 6in
           {\begin{center}
        \parbox{6in}{\footnotesize\baselineskip=12pt Fig.~\thefigure. #1}
            \end{center}}
        \else
             {\begin{center}
             {\footnotesize Fig.~\thefigure. #1}
              \end{center}}
        \fi}

\newcommand{\tcaption}[1]{
        \refstepcounter{table}
        \setbox\@tempboxa = \hbox{\footnotesize Table~\thetable. #1}
        \ifdim \wd\@tempboxa > 6in
           {\begin{center}
        \parbox{6in}{\footnotesize\baselineskip=12pt Table~\thetable. #1}
            \end{center}}
        \else
             {\begin{center}
             {\footnotesize Table~\thetable. #1}
              \end{center}}
        \fi}

\def\@citex[#1]#2{\if@filesw\immediate\write\@auxout
	{\string\citation{#2}}\fi
\def\@citea{}\@cite{\@for\@citeb:=#2\do
	{\@citea\def\@citea{,}\@ifundefined
	{b@\@citeb}{{\bf ?}\@warning
	{Citation `\@citeb' on page \thepage \space undefined}}
	{\csname b@\@citeb\endcsname}}}{#1}}

\newif\if@cghi
\def\cite{\@cghitrue\@ifnextchar [{\@tempswatrue
	\@citex}{\@tempswafalse\@citex[]}}
\def\citelow{\@cghifalse\@ifnextchar [{\@tempswatrue
	\@citex}{\@tempswafalse\@citex[]}}
\def\@cite#1#2{{$\null^{#1}$\if@tempswa\typeout
	{IJCGA warning: optional citation argument
	ignored: `#2'} \fi}}

 1
 1
 1

\font\ninerm=cmr9

\begin{document}

\newcommand{\st}{\scriptstyle}
\newcommand{\sst}{\scriptscriptstyle}
\newcommand{\mco}{\multicolumn}
\newcommand{\epp}{\epsilon^{\prime}}
\newcommand{\vep}{\varepsilon}
\newcommand{\ra}{\rightarrow}
\newcommand{\ppg}{\pi^+\pi^-\gamma}
\newcommand{\vp}{{\bf p}}
\newcommand{\ko}{K^0}
\newcommand{\kb}{\bar{K^0}}
\newcommand{\al}{\alpha}
\newcommand{\ab}{\bar{\alpha}}
\def\be{\begin{equation}}
\def\ee{\end{equation}}
\def\bea{\begin{eqnarray}}
\def\eea{\end{eqnarray}}
\def\li#1{\hbox{${}^{#1}$Li}}
\def\he#1{\hbox{$^{#1}{\rm He}$}}
\def\la{~\mbox{\raisebox{-.6ex}{$\stackrel{<}{\sim}$}}~}
\def\ga{~\mbox{\raisebox{-.6ex}{$\stackrel{>}{\sim}$}}~}
\def\CPbar{\hbox{{\rm CP}\hskip-1.80em{/}}}

\rightline{UMN-TH-1328/95}
\rightline{February 1995}
\centerline{\normalsize\bf BIG BANG NUCLEOSYNTHESIS\footnote{To be published in
the
proceedings of Beyond the Standard Model IV,
Lake Tahoe, CA, December 13-18, 1994, eds. J. Gunion,
T. Han, and J. Ohnemus (World Scientific, Singapore).}}
\baselineskip=16pt

\centerline{\footnotesize KEITH A. OLIVE}
\baselineskip=13pt
\centerline{\footnotesize\it School of Physics and Astronomy, University
of Minnesota}
\baselineskip=12pt
\centerline{\footnotesize\it Minneapolis, MN 55455, USA}
\centerline{\footnotesize E-mail: Olive@mnhep.hep.umn.edu}
\vspace*{0.3cm}
\baselineskip=13pt

\vspace*{0.9cm}
\abstracts{The current status of big bang nucleosynthesis and its implications
for physics beyond the standard model is reviewed. In particular, limits
on the effective number of neutrino flavors and extra Z gauge boson masses
are updated.}

\normalsize\baselineskip=15pt
\setcounter{footnote}{0}
\renewcommand{\thefootnote}{\alph{footnote}}

\section{Introduction}
The overall status of big bang nucleosynthesis is determined by the comparison
of the rather slowly changing theoretical predictions of the light element
abundances and the sometimes quickly changing observationally determined
abundances. The observed elements, D, \he3, \he4, \li7, have abundances
 relative  to hydrogen which span nearly nine orders of magnitude. By and
large,
these observations are consistent with the theoretical predictions and
play a key role in determining the consistency of what we refer to as
the standard big bang model and its extrapolation to time scales on the
order of one second.
Here, I will review the status of this consistency.  I will begin
by briefly outlining the key sequence of events in the early Universe
which leads to the formation of the light elements. I will then discuss
the current status of the observations in relation to theory of each of the
light elements.  Finally, I will  discuss the current limits on physics beyond
the standard model.

\section{A Brief Primer on the Theoretical Predictions}
Conditions for the synthesis of the light elements were attained in the
early Universe at temperatures  $T \la $ 1 MeV, corresponding to an age of
about 1 second.  At somewhat higher temperatures, weak interaction rates were
in equilibrium, thus fixing the ratio of number densities of neutrons to
protons. At $T \gg 1$ MeV, $(n/p) \simeq 1$.  As the temperature fell and
approached the point where the weak interaction rates were no longer fast
enough
to maintain equilibrium, the neutron to proton ratio was given approximately by
the Boltzmann factor, $(n/p) \simeq e^{-\Delta m/T}$, where $\Delta m$
is the neutron-proton mass difference. The final abundance of \he4 is very
sensitive to the $(n/p)$ ratio.

The nucleosynthesis chain begins with the formation of deuterium.
However, because the large number of photons relative to nucleons,
$\eta^{-1} = n_\gamma/n_B \sim 10^{10}$, deuterium production is delayed past
the point where the temperature has fallen below the deuterium binding energy,
$E_B = 2.2$ MeV.  When the quantity $\eta^{-1} {\rm exp}(-E_B/T) \sim 1$,
the nuclear chain begins at a temperature $T \sim 0.1 MeV$.

The dominant product of big bang nucleosynthesis is \he4 resulting in an
abundance of close to 25\% by mass. In the standard model,
 the \he4 mass fraction
depends primarily on the baryon to photon ratio,
$\eta$.  The change due to the uncertainty in the neutron half-life is small
(this effect is shown in Figure 1). When we go beyond the standard model, the
\he4 abundance is very sensitive to changes in the expansion rate which
can be related to the effective number of neutrino flavors as will
be discussed below. Lesser amounts of the other light elements are produced:
D and \he3 at the level of about $10^{-5}$ by number, and \li7 at the level of
$10^{-10}$ by number.

\vskip 1in

\fcaption{{The light element abundances from big bang
nucleosynthesis.}}
\vskip .2in

The resulting abundances of the light elements are shown in Figure 1 from the
calculations in ref. 1. The curves for the \he4 mass fraction, $Y$, bracket the
computed range based on the uncertainty of the neutron mean-life which
has been taken as $\tau_n = 887 \pm 2$ s. The \he4 curves have been adjusted
according to the corrections in ref. 2. Uncertainties in the produced \li7
abundances have been adopted from the results in ref. 3. Uncertainties in D and
\he3 production are negligible on the scale of this figure. The  boxes
correspond
to the observed abundances and will be discussed below.  It is clear that
as the observational boxes line up on top of each other, there is an
overall agreement between theory and observations in the range
$\eta_{10} = 10^{10} \eta  = 2.8$ -- 3.9.

\section{The Observations}
Because helium is produced in stars, it is very difficult to extract the
primordial abundance from the observations.  Ideally, one would want to look
for
primordial helium in regions where the stellar processing is minimal, i.e., in
regions where the abundances of elements such as carbon, nitrogen and oxygen
are
very low.  The \he4 abundance in very low metallicity regions is best
determined
from observations in extragalactic HII regions of HeII $\rightarrow$ HeI
recombination lines.  There are extensive compilations of observed
abundances of \he4, N, and O, in many different galaxies \cite{p,evan}. In
Figure 2, the \he4 vs. O data is shown along with its associated linear fit
\cite{OSt} (details of this fit are given in the second line of Table 1).

\vskip 1in

\fcaption{{The observed abundances of \he4 vs. O/H in
extragalactic HII regions along with a linear fit to the data.}}
\vskip .2in

In Table 1, various fits to the data and subsets of the data are given.
Details concerning the subsets of the data shown can be found in ref. 6. As one
can see there is a considerable degree of stability in these fits, leading to a
2 $\sigma$ upper limit of 0.238 for the primordial abundance of \he4. There is
in addition an overall systematic uncertainty of about 0.005 in $Y_p$ giving a
range (2 $\sigma$ plus systematic) of 0.221 -- 0.243 for $Y_p$ and is shown in
Fig. 1 as the large box bracketing the \he4 curves.

\begin{table}\begin{center}
\tcaption{Linear Fits for $Y$ vs. $O/H$.}\label{tab:smtab}

\begin{tabular}{|c|c|c|c|c|c|c|} \hline\hline
Set &  \# Regions & $r$ & ${\chi}^2/dof$ &  $Y_P$  &  $10^{-2}
 \times$ slope &$Y_P^{2\sigma}$ \\ \hline

All & 49 & 0.56 & 0.78 & $.234 \pm .003$ & $ 1.14 \pm 0.24$ & 0.239 \\
1st cut & 41 & 0.51 & 0.61 & $.232 \pm .003$ & $ 1.38 \pm 0.36$& 0.238 \\
-outliers & 34 & 0.45 & 0.70 &$.232 \pm .003$ & $1.39 \pm 0.38$& 0.238 \\
2nd cut & 21 & 0.41 & 0.64 & $.229 \pm .005$ & $2.37 \pm 1.13$ & 0.238 \\
-outliers & 19 & 0.40 & 0.70 & $.229 \pm .005$ & $2.42 \pm 1.15$ & 0.238\\
C & 22 & 0.35 & 0.71 & $.232 \pm .003$ & $1.58 \pm 0.54$ & 0.238 \\
\hline\hline
\end{tabular}
\end{center}
\end{table}

 It is more difficult to
compare the primordial deuterium  and \he3 abundances
with the observations. Despite the fact that
all observed deuterium is primordial,  deuterium is destroyed in stars.
A comparison between the predictions
of the standard model and observed solar and interstellar values of deuterium
must be made in conjunction with models of galactic chemical evolution.
The problem concerning \he3 is even more difficult.  Not only
is primordial \he3 destroyed in stars but it is very likely that low mass stars
are net producers of \he3. Thus the comparison between theory and observations
is
complicated not only by our lack of understanding regarding chemical evolution
but also by the uncertainties of the production of \he3  in stars.

It appears that D/H has decreased
over the age of the galaxy.  The (pre)-solar system abundance of deuterium
is\cite{geiss} D/H $\approx (2.6 \pm 1.0) \times 10^{-5}$,
corresponding to an age $t \sim 9$ Gyr, while the present ($t \sim 14$ Gyr) ISM
abundance of D/H is\cite{linsky} $1.65 \times 10^{-5}$. Thus, if $\eta_{10}$
is in the range 2.8 -- 3.9, then the primordial abundance of D/H is between 4.5
-- 8 $\times 10^{-5}$, and it would appear that significant destruction of
deuterium is necessary. Chemical evolution models
which destroy D by as much as a factor of 5 have been considered
recently\cite{sco}.

 Note that  there is a reported detection of D in a high redshift,
low metallicity quasar absorption system\cite{quas}
 with an abundance which may be the primordial one.
This observation is shown in Fig. 1 by the small box on the D/H curve at a
value of $\eta_{10} \approx 1.5$.  As one can see the corresponding value of
$Y_p$ (at the same value of $\eta$) is in excellent agreement with the data.
\li7 is also acceptable at this value as well. Due to the still some what
preliminary status of this observation (in fact a recent report\cite{t} claims
a
much lower D abundance along a different line of sight) and  the fact that it
can
also be interpreted as a H detection  in which the absorber is displaced in
velocity by 80 km s$^{-1}$ with respect to the quasar\cite{s}, it is premature
to fix the primordial abundance with that value. A high value
for the D abundance would require an even greater degree of D destruction over
the age of the galaxy.

There are however potential problems for \he3. The lower limit on $\eta$ was
derived\cite{ytsso} by noting that although stars can destroy \he3, even
very massive stars, still preserve at least 25 \% of the initial D + \he3.
(It is the sum of D and \he3 that is important as D is burned to \he3 in the
premain-sequence phase of stars.)  A value of $\eta_{10}$ lower than 2.8 would
yield (D+\he3)/H $> 10^{-4}$ so that even if the maximal amount of \he3
is destroyed, it would still exceed the presolar value\cite{geiss} of
(D+\he3)/H
$\approx 4.1 \pm 1.0 \times 10^{-5}$. But, in low mass stars, \he3 is produced
rather than destroyed.  A calculation\cite{it} of the final \he3 abundance
relative to the initial abundance of D + \he3 gives,
\be
(^3{\rm He/H})_f = 1.8 \times 10^{-4}\left({M_\odot \over M}\right)^2
+ 0.7\left[({\rm D+~^3He)/H}\right]_i
\ee
so that a 1 M$_\odot$ star produces 2.7 times as much \he3 as the initial D +
\he3 (for (D + \he3)$_{\rm initial} = 9 \times 10^{-5}$). This would lead to an
evolutionary behavior\cite{orstv} of the type shown in Fig. 3. The chemical
evolution model has been chosen so that
 D/H agrees with the data and assumes that $\eta_{10} = 3$.
The problem being emphasized concerns \he3 and can be seen by comparing
the solid curve with the filled diamonds.

\vskip 1in

\fcaption{{The evolution of
D/H (dashed curve), \he3/H (solid curve) and (D + \he3)/H (dotted curve)
  as a function of time.  Also shown are the data at the solar epoch
$t \approx 9.6$ Gyr and today for
D/H (open squares), \he3/H (filled diamonds) and
 (D + \he3)/H (open circle).\cite{orstv}.}}
\vskip .2in

A number of questions regarding the \he3 discrepancy can be raised. First, one
can ask whether or not the formula for
\he3 production is valid.  There is indeed evidence that \he3 is produced in
planetary nebulae\cite{rbw} where abundances are found as high as \he3/H $\sim
10^{-3}$. Secondly, one may ask how uniform are the \he3 measurements. As it
turns out, they are in fact not very uniform\cite{bbbrw} and show variations as
large as a factor of 5 between different HII regions.  Indeed, there may even
be a correlation between the size (mass) of the region and the amount of \he3
observed\cite{orstv}. Thus it may be possible that the solar values are
depleted in \he3. It is hoped that future \he3 observations will help to
resolve this puzzle.

Finally I turn to \li7.  Over the last several years, there has been a
considerable increase in the number of \li7 observations\cite{li}.  \li7 in
old,
hot, population II stars, is found to have a very nearly  uniform abundance.
For
stars with a surface temperature $T > 5500$ K and a metallicity than about
1/20th solar, the  \li7 abundance shows little or no dispersion beyond what is
consistent with errors of individual measurements. The corresponding mean \li7
abundance is Li/H =
$1.2
\pm .1
\times 10^{-10}$. Systematic errors, however, dominate the uncertainty in the
\li7 abundance, allowing a value Li/H $\la 2 \times 10^{-10}$. These values
(and
their uniformity) should be compared with observations of Li in younger stars
where the abundance can be much larger (by an order of magnitude) and shows
considerable dispersion. Two key questions remain, however: how much of
the observed Li is primordial (since Li is known to be produced); and how much
of the primordial Li remains in the stars where Li is observed?

Aside from the big bang, Li is produced together with Be and B in cosmic ray
spallation of C,N,O by protons and $\alpha$-particles.  Li is also produced by
 $\alpha-\alpha$ fusion.  Be and B have recently been observed in these same
pop II stars and in particular there are a dozen or so, stars in which both
Be and \li7 have observed.  Thus Be (and B though there is still a paucity of
data) can be used as a consistency check on primordial Li \cite{check}. Based
on
the Be abundance found in these stars, one can conclude that no more than 10-20
\% of the \li7 is due to cosmic ray nucleosynthesis leaving the remainder
(an abundance near $10^{-10}$) as primordial. It is also possible however, that
some of the initial Li in these stars has been depleted.  Standard stellar
models\cite{del} predict that any depletion of \li7 would be accompanied by a
very severe depletion of \li6.  Until recently, \li6 had never been observed in
hot pop II stars. The observation\cite{li6o} of \li6 which turns out to be
consistent with its origin in cosmic-ray nucleosynthesis and with a small
amount
of depletion as expected from standard stellar models is  a good indication
that
\li7 has not been destroyed in these stars\cite{li6}.

Consistency of the standard model of big bang nucleosynthesis relies on the
concordance between theory and observation of the light element abundances for
a single value of $\eta$. I now summarize the constraints on $\eta$ from each
of the light elements.   From the \he4 mass fraction, $Y < 0.238 (0.243)$,
we have that $\eta_{10} < 2.5 (3.9)$ as a $2 \sigma$ upper limit (the higher
value takes into account possible systematic errors).  Because of the
sensitivity to the assumed upper limit on $Y_p$, the upper limit on $\eta$
from D/H, though weaker is still of value.  From D/H $> 1.5 \times 10^{-5}$
we have $\eta_{10} < 7$.  The lower limit on $\eta$, comes from the upper
limit on D + \he3 and is $\eta_{10} > 2.8$ if one ignores \he3 production.
Finally, \li7  allows a broad range for $\eta$ consistent with other light
elements.  When both uncertainties in the reaction rates and systematic
uncertainties in the observed abundances are taken into account, \li7 allows
values of $\eta_{10}$ between 1.3 -- 4.5.  Taken all together, these bound on
$\eta$ constrain the fraction of critical density in baryons to be between
0.01 -- .1, for a hubble parameter, $h_o$, between 0.4 -- 1.0 (the
corresponding range for
$\Omega_B h_o^2$  is 0.010 -- 0.016).

\section{Constraints on Physics Beyond the Standard Model}
Limits on particle physics beyond the standard model are mostly sensitive to
the bounds imposed on the \he4 abundance. As is well known, the $^4$He
abundance
is predominantly determined by the neutron-to-proton ratio just prior to
nucleosynthesis and is easily estimated assuming that all neutrons are
incorporated into $^4$He,
\be
Y_p \approx {2 (n/p) \over 1 + (n/p)}
\ee
As discussed earlier, the neutron-to-proton
ratio is fixed by its equilibrium value at the freeze-out of
the weak interaction rates at a temperature $T_f \sim 1$ MeV modulo the
occasional free neutron decay.  Furthermore, freeze-out is determined by the
competition between the weak interaction rates and the expansion rate of the
Universe
\be
{G_F}^2 {T_f}^5 \sim \Gamma_{\rm wk}(T_f) = H(T_f) \sim \sqrt{G_N N} {T_f}^2
\label{comp}
\ee
where $N$ counts the total (equivalent) number of relativistic particle
species.
 The presence
of additional neutrino flavors (or any other relativistic species) at
the time of nucleosynthesis increases the overall energy density
of the Universe and hence the expansion rate leading to a larger
value of $T_f$, $(n/p)$, and ultimately $Y_p$.  Because of the
form of eq. (\ref{comp}) it is clear that just as one can place
limits\cite{ssg} on $N$, any changes in the weak or gravitational coupling
constants can be similarly constrained (for a recent discussion see ref. 23.

In the standard model, the number of particle species entering into eq.
(\ref{comp}) can be written as $N  = 5.5 + {7 \over 4}N_\nu$ (5.5 accounts for
photons and $e^{\pm}$). The observationally derived primordial \he4
abundance\cite{OSt} of
$Y_p = 0.232 \pm 0.003 \pm .005$ translates into a best value for $N_\nu = 2.2
\pm 0.27 \pm .42$ which implies a $2 \sigma $ upper limit of 2.74 which is
extended to $N_\nu < 3.16$ when systematics are included.
At face value, such a limit would exclude even a single additional scalar
degree of freedom (which counts as ${4 \over 7}$) such as a majoron unless it
decoupled early enough\cite{oss} so that its temperature, $T_B$ at the time of
nucleosynthesis was suppressed and $(T_B/T_\nu)^4 < {7 \over 4}(.16) = .28$.
In models with right-handed interactions, and three right-handed neutrinos, the
constraint is more severe. The right-handed states must have decoupled early
enough to ensure $(T_{\nu_R}/T_{\nu_L})^4 < (.16)/3 \simeq .05$. The
temperature of a decoupled state is easily determined from entropy
conservation, $(T_x/T_\nu) = \left( (43/4)/N(T_d) \right)^{1/3}$. One
additional scalar requires $N(T_d) > 28$ or decoupling must have occurred
above the QCD phase transition at a temperature $T_d > T_{\rm QCD} \sim 200$
MeV.  Three right-handed neutrinos would require $N(T_d) \ga 100$, so that $T_d
> M_W$. If right-handed interactions are mediated by additional gauge
interactions, the associated mass scale becomes $M_{Z'} > O(100)$ TeV!

The limits on $N_\nu$, however,  are sensitive to the upper limit on \he4 which
is in turn sensitive to assumed systematic errors and to the lower bound on
$\eta$.  In addition, the limits described above may be overly
restrictive\cite{lang}. The best value for $N_\nu$ is 2.2 and may in fact be
unphysical if $\nu_\tau$ is lighter than $\sim 1$ MeV, as is quite likely.  In
this case, the limits on $N_\nu$ must be accordingly renormalized\cite{ost2}.
In Fig. 4, the effect of renormalizing the limit on $N_\nu$ is shown.

\vskip 1in

\fcaption{{The 95 \% CL upper limit on $N_\nu$ as a function of
the systematic uncertainty in $Y_p$. The solid (dashed) curves correspond to
the condition that $N_\nu>$ 2 (3). Each of these two cases is shown for three
choices of $\eta_{10}$: 1.5, 2.8, and 4.0. The smaller values of $\eta$
yield weaker upper limits and systematic errors have been
described by a top-hat distribution\cite{ost2}.}}

In summary, I have argued for the overall agreement between theory and
observations as they pertain to the light element abundances as well as
concordance between big bang nucleosynthesis and galactic cosmic-ray
nucleosynthesis.  There are however, open issues: Are the quasar line-of-sight
measurements\cite{quas} of D/H real; Why isn't there more \he3, particularly in
the solar system; Can the statistical and systematic errors in \he4
measurements be reduced; Can the large systematic errors in the \li7 abundance
be reduced?  Clearly new data will be necessary to resolve these problems.
Nevertheless, in spite of these uncertainties, nucleosynthesis is still able to
set strong constraints on physics beyond the standard model.

This work was supported in part by  DOE grant DE-FG02-94ER40823.

\medskip

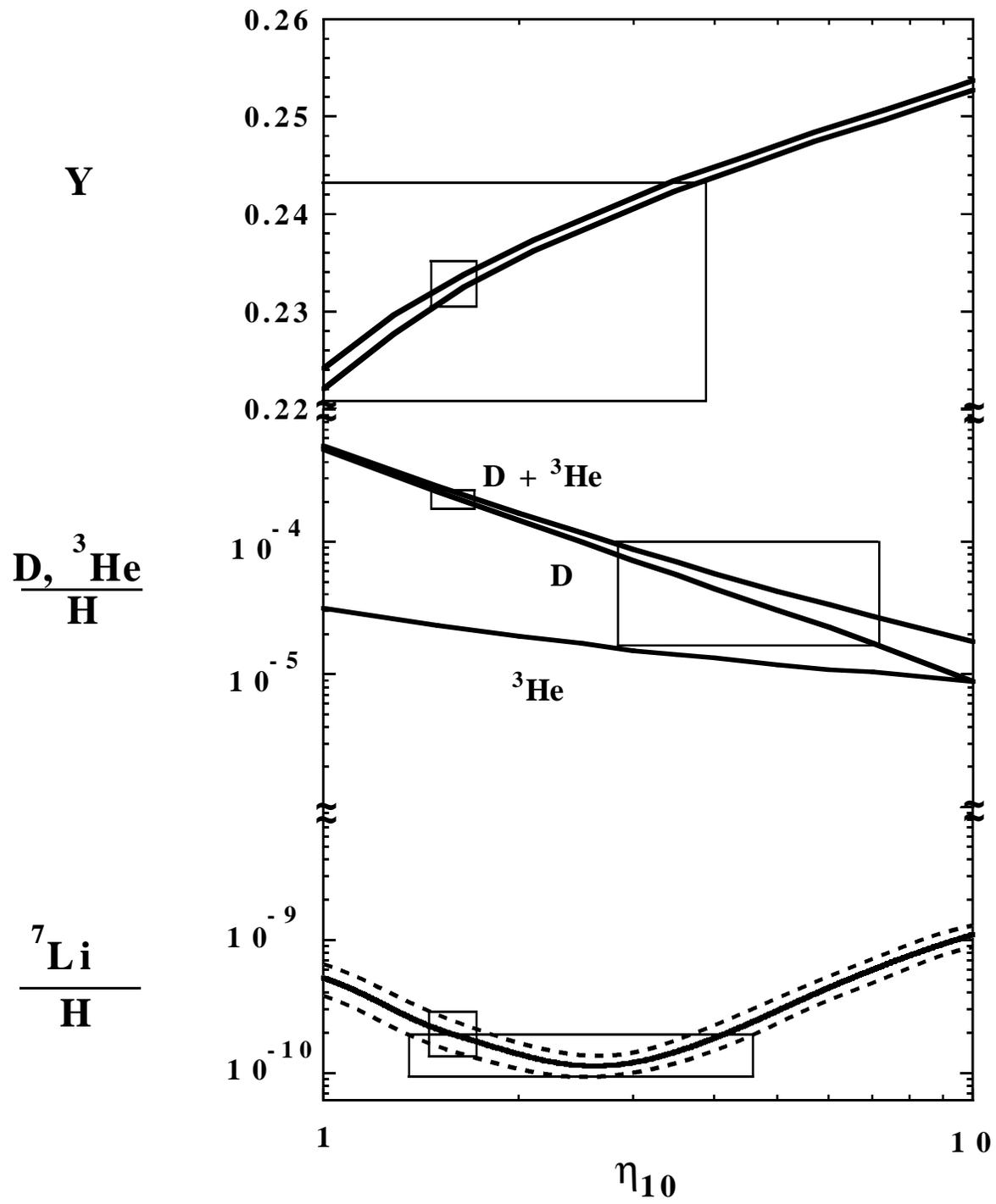
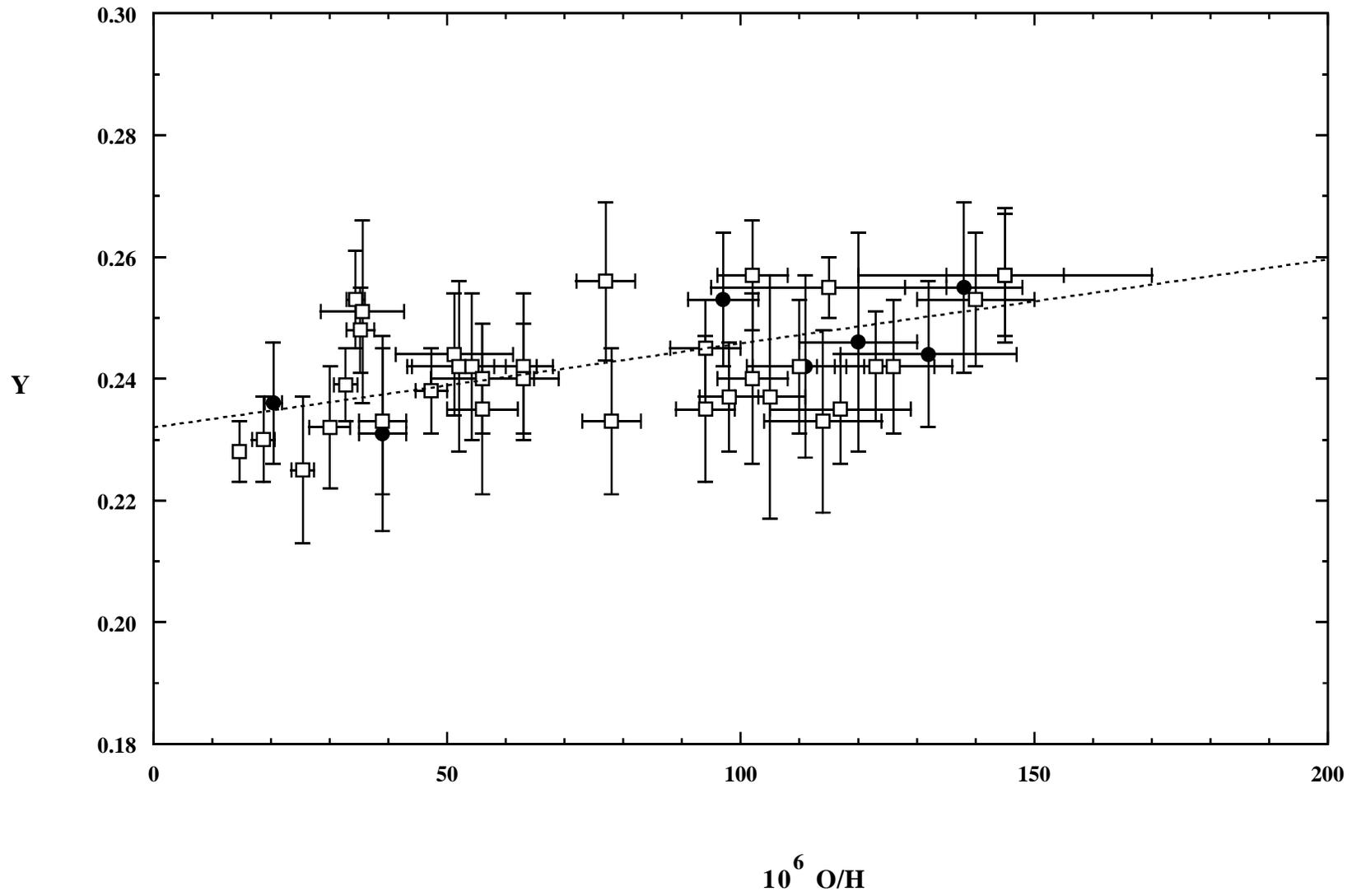
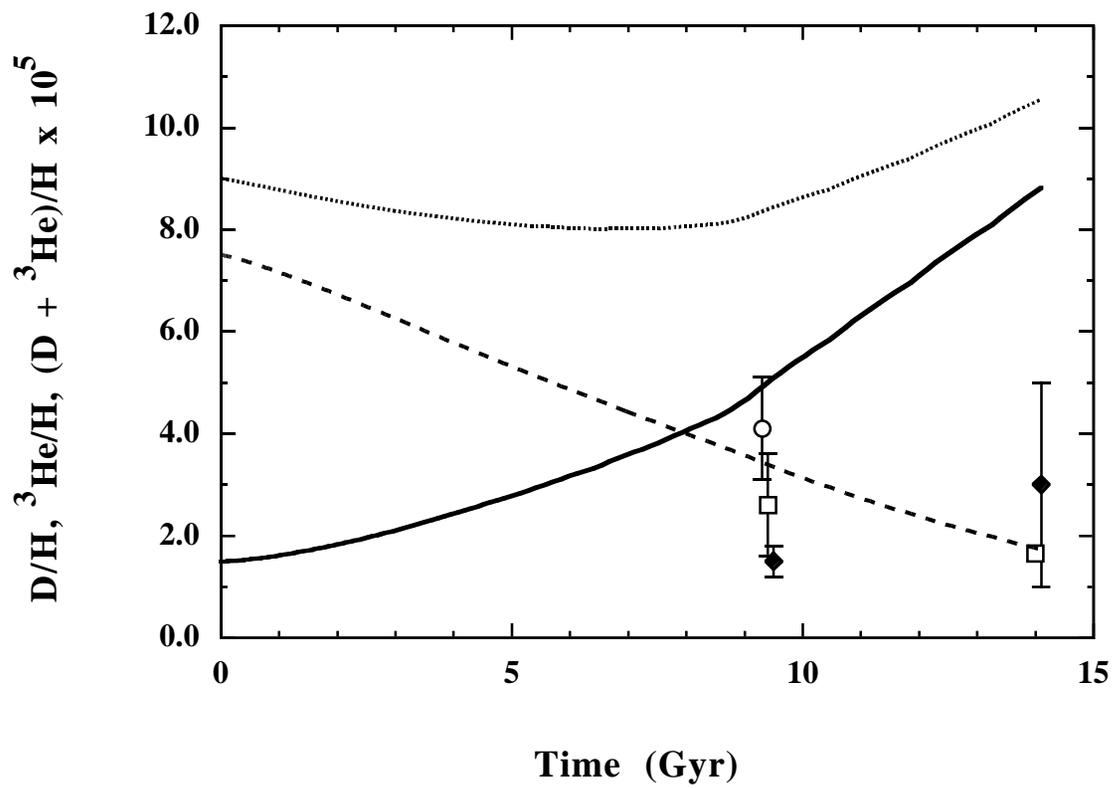
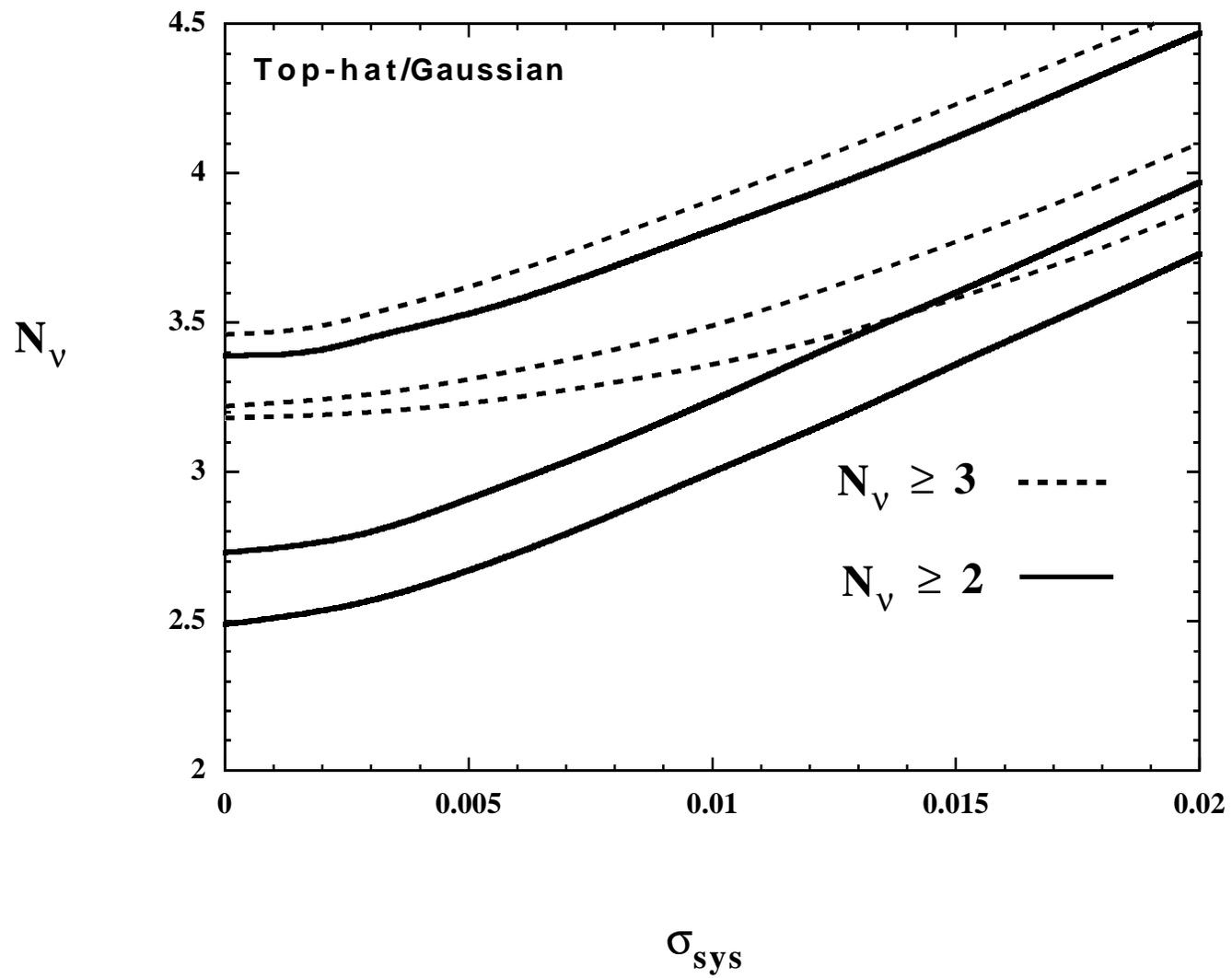
\end{document}